\documentstyle[12pt]{article}

\begin{document}
\begin{titlepage}
%\vspace{1cm}
{\hskip 9.5cm} CBPF-NF-075/95 \\

%\vspace{1cm}
\centerline{\large \bf
THE DISSIPATIVE POTENTIAL INDUCED BY QCD}
%\vspace{0.4cm}

\centerline{\large \bf AT FINITE TEMPERATURE AND DENSITY
\footnote{This work is partly supported by the National Science
Foundation of China (NSFC). }}

\vspace{0.5cm}
\centerline{\bf Wu-Sheng Dai $^{1,2}$, Song Gao $^{3}$,
Xin-Heng Guo $^{3,4}$, Xue-Qian Li
$^{1,2}$,}

%\vspace{0.8cm}

\centerline{\bf  Peng-Nian Shen $^{1,3}$ and Ru-Keng Su $^{5}$}

%\vspace{0.8cm}
{\small
\begin{center}
1.   China Center of Advanced Science and Technology,
  (World Laboratory), Beijing, China.
%\vspace{8pt}

2. Department of Physics, Nankai University, Tianjin, China.
%\vspace{8pt}

3. Institute of High Energy Physics, Beijing, China.
%\vspace{8pt}

4. Centro Brasileiro de Pesquisas Fisicas, Rio de Janeiro, Brasil.

%\vspace{8pt}

5. Department of Physics, Fudan University, Shanghai, China.

\end{center}
}
%\vspace{1.5cm}

\centerline{\large \bf Abstract}
%\vspace{0.8cm}

In the framework
of QCD at finite temperature we have obtained dissipative terms
for the effective potential between $q$ and $\bar q$ which would partly
explain the $J/\psi$ suppression in the Quark Gluon Plasma
(QGP). The derivation of the 
dissipative potential for QGP is presented and the case for
Hadron Matter (HM) is briefly
discussed. The suppression effects are estimated based on  simple
approximations.\\

%\vspace{0.5cm}

\end{titlepage}

\baselineskip 22pt

\noindent 1. Introduction

\vspace{0.5cm}

It was supposed that a suppression of $J/\Psi$ production in Relativistic
Heavy Ion Collisions (RHIC)
could be a clear signal for formation of a new matter
phase, e.g. the Quark Gluon Plasma (QGP) \cite{Matsui}, later the phenomenon
was observed by the NA38 collaboration \cite{NA38}. However, it is too
early to celebrate the discovery of QGP, because rigorous studies indicate
that in the Hadron Matter (HM) phase,  the suppression phenomenon also
exists. Thus many authors study the mechanisms which cause the suppression of
$J/\Psi$ production in the QGP and HM phases separately \cite{Matsui}
\cite{Capella}. They investigate how the mass of $J/\Psi$ shifts with
temperature and density of the QGP and HM medium, especially the dependence
of the quark and gluon condensates on temperature and density are introduced
to explain the mass shifts and then the suppression of $J/\Psi$ production.

The RHIC experiments provide an atmosphere of hot and dense matter state,
no matter it is in QGP or HM phases. The two situations are quite
different in many aspects and an important research subject is to look
for one or several clear signals to confirm or negate formation of QGP.
The $J/\Psi$ suppression is one of the candidates, but obviously, 
observation of such suppression is not enough for drawing a
definite conclusion, one should investigate
the details of the  mechanisms which result in the suppression in QGP and HM.
If one starts with a fundamental principle, i.e. QCD,
instead of phenomenologically obtained potentials, he may expect to derive
different potential forms for QGP and HM, and it just is the task of this
work.

The effective potential between quark-antiquark or quark-quark
can be written as
\begin{equation}
V(r)=-{\alpha_{eff}\over r}+\sigma r.
\end{equation}
In the regular zero-temperature theory $\alpha_{eff}=4\alpha_s/3$ for
meson and $2\alpha_s/3$ for baryon where $\alpha_s=g_s^2/4\pi$
and $g_s$ is the QCD coupling constant. At finite temperature, the confinement
constant becomes temperature-dependent \cite{Hashimoto},
\begin{equation}
\sigma (T)=\sigma (0)[{T_{dec}-T\over T_{dec}}]^{\delta}\theta(T_{dec}-T),
\end{equation}
where $T_{dec}$ is the deconfinement temperature and $\delta$ is an uncertain
parameter. Above $T_{dec}$, the linear confinement potential disappears, but
it does not mean that the bound state dissolves, because
a binding energy can also be provided by the Coulomb-type potential. By analysis
of the medium state, a Debye screening mechanism is suggested as
\begin{equation}
V(r)=-{\alpha_{eff}\over r}exp(-{r\over r_D}),
\end{equation}
where $r_D$ is the Debye screening length. It is a function of temperature
\cite{Matsui} and  may be obtained by models. The dissolution condition
can be written as
$$r_D\leq 0.84 r_B=1.68/(\alpha_{eff}m_c). $$

Seeking for other possible mechanisms which may cause the $J/\Psi$ suppression,
we turn to re-study the situation. We will start with the QCD theory
and its low energy phenomena at finite temperature and density.

Let us briefly retrospect how one derives the Coulomb-type
potential from the field theory.
Considering a t-channel scattering between two quarks or quark-antiquark,
the amplitude in momentum space is
\begin{equation}
\label{parrot}
M=\overline u_1(\vec p_1)\gamma_{\mu}u_2(\vec p_2)G^{\mu\nu}(k^2)
\overline u_3(\vec p_3)\gamma_{\nu}u_4(\vec p_4),
\end{equation}
where $G^{\mu\nu}(k^2)$ is the full gluon propagator and
for $q-\bar q$ interaction $u_3$ and $u_4$ become $v_3$ and $v_4$.
There is also an extra s-channel annihilation diagram for $q-\bar q$, where
$u_2$ and $u_3$ turn to $v_2$ and $v_3$. Setting $k_0=0$, a
Fourier transformation of the amplitude gives rise to
the effective potential in
configuration space. Generally speaking, in $t-$channel the
momentum transfer is space-like, i.e. $k^2\leq 0$.

It is known that the photon and gluon propagators cannot provide an
imaginary part, even at finite temperature and density in the common
sense. The reason is that an absorptive part of the photon and gluon
propagators corresponds to on-shell photon or gluon (gauge bosons), i.e.
$k^2=0$, it indeed manifests an emission or absorption of a real photon or
gluon from the quark (antiquark). At s-channel, the intermediate gauge bosons
have momentum $k^2=s=(p_{q}+p_{\bar q})^2\geq 4m_q^2>0$, thus the $k^2=0$
condition cannot be satisfied, unless via loops. Whereas at t-channel
the momentum transfer carried by the gauge boson is $k^2=(p_q-p_q')^2\leq 0$,
so it seems that an imaginary part can appear. But it is not true, because
only at the boundary point $(p_q-p_q')^2=0$, it means that  if at
both sides of the interaction, the quark is on its mass shell
$E_q=E_q'$ and $\vec p_q=\vec p_q'$,
i.e. the $\theta=0$ forward scattering. For the Fourier transformation
the integration should be restrained in a momentum conservation
allowed region, so for the exact on-shell situation, the integration region is
zero (see below in the context for details), namely $\int_{0}^{0}|\vec k|f(k)
$ would result in a null contribution to the absorptive part. By contraries,
for the real part of the propagator which gives rise to
the regular potential, $k^2=0$ condition is not imposed, so
the integration region does not have any restriction (see below). That is why
in the regular theories, photon and gluon propagators do not contribute
an imaginary (absorptive) part, but only a real (dispersive) part.

However, in a bound state under consideration, the situation is different.
If the heavy quark or antiquark ($c$, $\bar c$) is on-shell, it can be
described as
\begin{equation}
p_{\mu}=m_cv_{\mu},
\end{equation}
where $v_{\mu}$ is the four-velocity, but in a bound state there is an
integration region \cite{Bardeen}
\begin{equation}
\label{residue}
p_{\mu}=m_cv_{\mu}+k_{\mu},
\end{equation}
where $k_\mu$ is the "residue" momentum, and $|\vec k|\sim k_0\leq \Lambda
_{QCD}$. Thus the integration region allowed by the energy-momentum
conservation is no longer zero, but can be from zero to $\Lambda_{QCD}$
(or a smaller value than $\Lambda_{QCD}$). It can indeed contribute an
absorptive part which finally results in a dissipative term in the effective
potential. In the next section we will present the details.

Assuming the temperature and density are above $T_{dec}$ and $\rho_{dec}$,
the QGP phase is reached. Without the linear confinement term whose source
is purely non-perturbative and obscure so far,
we only need to take into account the one-gluon-exchange diagram
which results in the interaction between
 quarks and quark-antiquark. It is well known that at $T=0$, the leading
term of the one-gluon-exchange contribution is the familiar Coulomb-type
potential $-\alpha_{eff}/r$, but as the non-perturbative QCD effects
which are characterized by the non-vanishing quark and gluon condensates are
taken into account, the potential form is modified and corrections
related to $1/m_Q$ ($m_c$ or $m_b$) emerge \cite{Shen}\cite{Liu}.
When the temperature is non-zero and the density is above the regular one,
more extra contributions appear. In this
work, ignoring all spin and spin-orbit dependent terms and
under some simple approximations,  we obtain a new
Coulomb-type potential as
\begin{equation}
\label{dog}
V(r)=-{\alpha_{eff}\over r}[1+i(a+\alpha_s b)],
\end{equation}
where $a$ and $b$ are functions of temperature and density. This new
term turns the $J/\Psi$ charmonium into a dissipative system and a
dissolution is expected.

In the next section, we derive eq.(\ref{dog}) in the framework of QCD at
finite temperature and in Sec.III, we discuss its significance and the
situation for HM phase.\\

\noindent II. Formulation.

In the QGP atmosphere
the propagator of fermion at finite temperature and density in momentum
space can be written as \cite{Kuo}
\begin{equation}
S_F(k)=(\rlap /k+m)[{i\over k^2-m^2}-2\pi\delta (k^2-m^2)f_F(k\cdot u)],
\end{equation}
where  
\begin{equation}
f_F(x)={\theta (x)\over e^{\beta(x-\mu)}+1}+{\theta (-x)\over
e^{-\beta (x-\mu)}+1}.
\end{equation}
In the expression $\beta=1/T$, $u$ is the four-velocity of the medium,
generally $u=(1,\: \vec 0)$ in the laboratory frame, so $k\cdot u=k_0$,
$\mu$ is the chemical potential and is related to the density of the medium.
Whereas the propagator of gluon reads
\begin{equation}
D_{\mu\nu}(k)=[{i\over k^2}+2\pi\delta (k^2)f_B(k\cdot u)](-g_{\mu\nu}
+{k_{\mu}k_{\nu}\over k^2})
\end{equation}
where
\begin{equation}
f_B(k\cdot u)={1\over e^{\beta k_0}-1}.
\end{equation}
It is noted that we are working in the real-time scheme \cite{Chou}.

As aforementioned, if the charm-quark in $J/\Psi$ bound state is strictly
on mass shell, the t-channel scattering demands $k^2\leq 0$, so that
the fermion propagator does not have the temperature-dependent term because
the intermediate fermion of non-zero mass cannot be an on-shell real particle. However,
the soft gluon interaction can make the charm quark deviate from its mass
shell, namely the $k^2$ can be equal to or
greater than zero in eq.(\ref{residue})
and it represents an emission or absorption of real photon or gluon,
($k^2=0$ and $|\vec k|\sim\Lambda_{QCD}$), moreover a small
time-like momentum transfer
$k^2=m_q^2>0$ is also reasonable, then the intermediate fermion is an on-shell
real particle.

Omitting the loop corrections, the lowest order Feynman diagrams are
depicted in Fig.1. In general, the contribution can be decomposed into a form
\begin{equation}
f(\mu,T)=f_1 (T)+g^2f_2(T)+g^2f_3(T,\mu),
\end{equation}
where $f_1(T), f_2(T)$ and $f_3(T, \mu)$ are from one gluon exchange, gluon condensate and quark condensate diagrams respectively shown in Fig. 1. It is noted that only $f_3$ depends on the density of QGP, but not $f_1$ and $f_2$. Both $g^2 f_2(T)$ and $g^2 f_3 (T, \mu)$ belong to higher order expansion, but while discussing the density dependence, $f_3 (T, \mu)$ becomes the leading term, so we drop out $f_2 (T)$ but keep $f_3 (T, \mu)$ in our calculations.
The second term with
the $\delta-$functions in the propagators correspond to the on-mass-shell
intermediate particle, therefore turns to be a real particle,
so it can feel the influence from the surrounding atmosphere, namely the
density and temperature.

The traditional way for obtaining the potential is only to deal with the
off-shell part of the propagator, namely to set $k_0=0$ and carry out
a three-dimensional Fourier transformation of the two quark scattering
amplitude. All the details including the loop corrections and
non-perturbative effects are presented in literatures \cite{Shen}\cite{Liu}
\cite{Landau}\cite{Gupta}, instead this work is only focused on the temperature
and density-dependent parts which would contribute  additional
modification terms to the potential, we denote them as $V^T_{G}(r)$ and
$V^T_q(r)$ corresponding to Fig.1 (a) and (b) respectively.

From Fig.1 (a), one has
\begin{equation}
\label{cat}
V^T_{G}(r)={1\over M_{J/\Psi}}({-16\pi i\over 3}\alpha_s)(-2\pi)
\int{d^4k\over (2\pi)^4}\delta (k^2)e^{-ik\cdot (x-y)}{1\over e^{\beta k_0}
-1},
\end{equation}
where the factor $({-16\pi i\over 3}\alpha_s)$ coming as a common factor
due to the color singlet condition of hadrons and it makes the leading
Coulomb term be $-{4\alpha_s\over 3r}$ as required. It is also noted that
here one does not need to set $k_0=0$ as in ref.\cite{Landau} and the final
integration region for $|\vec k|$ is from 0 to $\Lambda_{QCD}$ as discussed
in section I. The factor ${1\over
M_{J/\Psi}}$ guarantees the dimension of $V_G(r)$ right and comparing with
the traditional way for deriving the potential, there is an extra integral
over $k_0$, so a factor ${1\over E}\approx {1\over M_{J/\Psi}}$
is needed. This is equivalent to multiply a time factor $\tau$ ($\tau\cdot
E\sim 1$).

Taking the spontaneous requirement $x_0=y_0$ (potential means an
instantaneous interaction) a straightforward calculation gives
\begin{eqnarray}
\label{lion}
V^T_{G}(r) &=& {-16\pi i\over 3}\alpha_s{1\over (2\pi)^2}\cdot {1\over rM_{J/\Psi}}
\sum_{n=1}^{\infty}{1\over n^2\beta^2+r^2}\times \nonumber \\
&& [r(1-e^{-n\beta\Lambda}\cos\Lambda r)-n\beta e^{-n\beta\Lambda}\sin\Lambda
r],
\end{eqnarray}
where $\Lambda\leq\Lambda_{QCD} $ is a parameter  and
in the expression we deliberately pull out the $1/r$ factor (see below).
It is a series which absolutely converges for any finite $r$ and $\beta$.
If $r$ is not very large (a few tenths of $fm$ in our case), one only needs
to take first a few terms for numerical computations. One can notice that if
$\Lambda\sim 0$, i.e. for quarks are exactly on mass-shell, $V^T_G(r)$
vanishes and at $T\rightarrow 0$, it is also zero, this is consistent
with our common knowledge.
To see its meaning and avoiding tedious calculation, setting
$\Lambda\rightarrow\infty$ we approximate
eq.(\ref{lion}) to an integral as
$$\sum_{n=1}^{\infty}{r\over n^2\beta^2+r^2}\sim \int_{1}^{\infty} dx
{r\over x^2\beta^2+r^2}={1\over\beta}({\pi\over 2}-arctan{\beta\over r}).$$
Thus the $V^T_{G}(r)$ is recast as
$$V^T_G(r)={-16\pi i\over 3}\alpha_s{1\over (2\pi)^2}{1\over\beta r}({\pi\over 2}
-arctan{\beta\over r}).$$
Below we will evaluate its contribution in terms of the series solution (\ref{cat}).

Similarly, for the quark contribution shown in Fig.1 (b), one has
\begin{eqnarray}
\label{tiger}
V^T_q(r) &=& {-32\pi i\over 9}\alpha_s^2{<\psi_q\overline\psi_q>_T\over m_q^3}\cdot
{1\over r}[{1-\cos\mu r\over r}-\sum_{n=1}^{\infty}(-1)^n
{1\over n^2\beta^2+r^2}\times \nonumber \\
&& [2r\cos\mu r-(r\cos\Lambda r+n\beta\sin\Lambda r)
e^{-n\beta \Lambda}(1+e^{\mu r})],
\end{eqnarray}
for $\Lambda\geq\mu$ and
\begin{eqnarray}
V^T_q(r)&=& {-32\pi i\over 9}\alpha_s^2{<\psi_q\overline\psi_q>_T\over m_q^3}\cdot
{1\over r}[{1-\cos\mu r\over r}-\sum_{n=1}^{\infty}(-1)^n
{e^{-n\beta\mu}\over n^2\beta^2+r^2}\times \nonumber \\
&& [r(2-(e^{n\beta\Lambda}-e^{-n\beta\Lambda})\cos\Lambda r)+
n\beta\sin\Lambda r(e^{n\beta\Lambda}-e^{-n\beta\Lambda})].
\end{eqnarray}
for $\Lambda<\mu$, it is obvious that as $\Lambda\rightarrow 0$, $V^T_q(r)$
vanishes.
While deriving eq.(\ref{tiger}), we approximate $E=\sqrt{\vec k^2+m_q^2}
\approx |\vec k|$ in the integral due to the smallness of $m_q$.
In the expression $<\psi_q\overline\psi_q>_T$ is the quark condensate
at finite temperature T and its significance will be discussed in the
next section. \\

\noindent III. Discussions.

The additional potential terms induced by QCD at finite temperature
and density are imaginary and tend to zero
as the temperature and density approach to zero. It makes sense because if
there is no hot and dense atmosphere for the gluon and quarks, the
additional terms do not exist at all, so this scenario
would not affect the regular lifetime (about $7.5\pm 0.4\times 10^{-21}$ sec.)
of the $J/\Psi$ which
mainly is determined by the s-channel annihilation process of $q\bar q$
into three gluons. When $T$ is very high (a few hundred MeV) and
$\rho\gg \rho_0$, the damping factor becomes substantial and it is the QGP
situation. (see below)

The temperature and density effects turn the Hamiltonian of $J/\Psi$
into complex, and the new Hamiltonian describes a dissipative quantum
system. Because it is dissipative, the system would dissolve after a
certain time, for example, into $D\overline D$ or
$D_s\overline D_s$ etc. Dissolution to $D\overline D$ or $D_s\overline D_s$
needs to absorb energies from atmosphere because $M_{J/\Psi}<2M_D$, in
QGP case, it is possible, whereas in the vacuum circumstance, due to the
kinematic constraint of the final state phase space, even there were a
dissipative term, the dissolution would not occur.

From eqs.(\ref{lion}, \ref{tiger}) one can notice that the coefficients of
$i$ in $V^T_G(r)$  and $V^T_q(r)$ are always negative,
This is a very important point because it makes the system dissipative.

Since $V^T_G(r)$ and $V^T_q(r)$ are very complicated functions of $r$, it would be
extremely difficult to solve the Schr\"{o}dinger equation with such a
potential. Therefore to see the physical significance, we would choose a
simple but reasonable approximation.

\noindent (a) Estimation of the dissipation effects.

(i) Looking at the additional Hamiltonian, one can note that they can be
written as $if_{q(G)}(r){1\over r}$ where $f_{q(G)}(r)$ are complicated
functions of $r$ and given in eqs.(\ref{lion}, \ref{tiger}). Our approximation
is to treat $f_{q(G)}(r)$ as average values instead of functions of $r$,
namely, we approximate $f_{q(G)}(r)$ as $f_{q(G)}(\bar r)$ where $\bar r$ is
the average radius of $J/\Psi$. Matsui \cite{Matsui} suggests
$0.2\leq \bar r_{_{J/\Psi}}\leq 0.5$ fm.

(ii) As $\sigma (T)$ disappears above the deconfinement temperature $T_{dec}$,
the stationary Schr\"{o}dinger equation becomes
\begin{equation}
\label{bird}
-{1\over 2m_{red}}\nabla^2\phi-{4\alpha_s\over 3r}[1+i(a+\alpha_s b)]\phi
=E\phi
\end{equation}
where $\phi$ is the wavefunction of $J/\Psi$, $m_{red}$ is the reduced mass
which equals $m_c/2$, and
\begin{eqnarray}
\label{dragon}
 a &=& {1\over\pi} \sum_{n=1}^{\infty}{1\over n^2\beta^2+\bar r^2}\times
 [\bar r(1-e^{-n\beta\Lambda}\cos\Lambda \bar r)-n\beta e^{-n\beta\Lambda}
\sin\Lambda\bar r]
\nonumber \\
&\stackrel{\Lambda\rightarrow\infty}{\longrightarrow}&
{1\over\pi}{1\over\beta}({\pi\over 2}-arctan{\beta\over \bar r})
\\
 b &=& \sum_{q=u,d,s}{8\over 3}{<\psi_q\overline\psi_q>\over m_q^3}
[{1-\cos\mu\bar r\over \bar r}-\sum_{n=1}^{\infty}(-1)^n
{1\over n^2\beta^2+\bar r^2}\times \nonumber \\
&& [2\bar r\cos\mu \bar r-(\bar r\cos\Lambda\bar r+n\beta\sin\Lambda\bar r)
e^{-n\beta \Lambda}(1+e^{\mu\bar r})], \;\;\;\;\;\;\;\;{\rm for}\; \Lambda\geq
\mu; \\
 b'&=& \sum_{q=u,d,s}{8\over 3}{<\psi_q\overline\psi_q>\over m_q^3}
[{1-\cos\mu\bar r\over \bar r}-\sum_{n=1}^{\infty}(-1)^n
{e^{-n\beta\mu}\over n^2\beta^2+\bar r^2}\times \nonumber \\
&& [\bar r(2-(e^{n\beta\Lambda}+e^{-n\beta\Lambda})cos\Lambda\bar r)
+n\beta\sin\Lambda\bar r(e^{n\beta\Lambda}-e^{-n\beta\Lambda})]
\;\;\;\;\;\;\;\;{\rm for}\; \Lambda<\mu.
\end{eqnarray}
The eigenenergy of eq.(\ref{bird}) is simple and reads
\begin{equation}
(E_j)_{eff}={-8m_{red}\over 9j^2}(\alpha_s)_{eff}^2,
\end{equation}
where $(\alpha_s)_{eff}=\alpha_s[1+i(a-\alpha_s b)]$. For the ground state
$j=1$.

(iii) The evolution of the quantum system can be expressed as
\begin{eqnarray}
\phi_{J/\Psi}(t) &=& \phi_{J/\Psi} (0)e^{-iE_{eff}t} \nonumber \\
&=& \phi_{J/\Psi} (0)exp(i{8m_{red}\alpha_s^2\over 9}[1-(a+\alpha_sb)^2]t)
\cdot exp({-16m_{red}\over 9}(a+\alpha_sb)\alpha_s^2t).
\end{eqnarray}
There exists a damping factor. In the calculations, we ignore the temperature
dependence of $\alpha_s$ \cite{Furnstahl}.

\noindent (b) Numerical evaluations.

(i) If the QCD expansion (including the condensates) converges, $|b\alpha_s|$
should be smaller than $a$. Typically, $\alpha_s\approx 0.3$ in the potential
model, the term $\alpha_sb$ can compete with the term $a$. $b$ is
proportional to the quark condensate at finite temperature
$<\psi_q\overline\psi_q>_T$ which decreases with the increase of temperature.
However, we will argue in the following that the ratio $\frac{<\psi_q\overline\psi_q>_T}{m_{q}^{3}}$ where $m_q$ is the constituent quark mass is almost independent of temperature. 

The constituet quark mass is defined as the pole of the quark propagator 
\begin{equation}
\label{pole}
\Sigma (p^2 =m_{q}^{2})=m_{q},
\end{equation}
where $\Sigma (p^2)$ is the dynamical mass related to quark condensate \cite{guo}
\begin{equation}
\label{pole1}
\Sigma (p^2) \sim \frac{1}{p^2}<\psi_q\overline\psi_q>\alpha_{s}(p^2).
\end{equation}
Hence the ratio $\frac{<\psi_q\overline\psi_q>_T}{m_{q}^{3}}$ depends only on $\alpha_{s}(m_{q}^{2})$ which dependence on temperature is ignored. Therefore, the parameter $b$ is almost temperature independent.

(ii) Thus we have
$$|\phi_{J/\Psi}(t)|^2=|\phi_{J/\Psi}(T=0)|^2e^{({-2t\over \tau_0})}.$$
A typical time factor is about $\tau_0\sim
7\times 10^{-22}$ sec., for $0.2\leq
\bar r\leq 0.5$ fm and here we use $\bar r=0.4$ fm. Even though this
numerical value cannot be taken very seriously, the order of magnitude is
reasonable.

The size of the collision region is about $10\sim 100$ fm, a particle
produced at the center of the region needs
$1\times 10^{-22}\sim 10^{-21}$ sec. to travel to the boundary,
so $\tau_0$ is of the same order as
the traverse time and definitely results in a suppression of $J/\Psi$
production (in fact a bulk decay) observed in experiments.

In fact, in such a small time interval, according to the regular theory
c and $\bar c$ hardly combine into a bound state, so it is equivalent to
the effective screening.

(iii) It is worth noting that only $b$ depends on the density via the
chemical potential $\mu$. From eq.(\ref{tiger}), numerical results show
that as $|\mu |$ gets larger the contribution of $V^T_q(r)$ becomes more
competitive to $V^T_G(r)$, but
by the common sense, $|b\alpha_s|<a$.

\noindent (c) For the hadron matter phase, gluons and quarks do not directly feel
the temperature and density of the hadron medium, therefore the gluon
propagator cannot be influenced by the medium. In this case the temperature
and density effects can appear in two ways. One is that the quark and gluon
condensates are modified in the medium but different from that in QGP,
\cite{Hatsuda}, while another way is via loops in s-channel.

Our results show that the temperature and density effects can cause a
suppression of $J/\Psi$ production, but the mechanism is different from
the traditional Debye screening effect, i.e. it contributes a dissipative part
to the potential. This additional term makes the quantum system dissolve
by a time scale $7\times 10^{-22}$ sec. This mechanism can only exist
in QGP phase but not in the HM phase.          \\

\vspace{1cm}

\noindent Acknowledgment:

We benefit greatly from fruitful discussions with
Profs. W.Q. Chao, C.S. Gao and B. Liu
on this subject, the author is indebted to Prof.Z. Lu who helped us
to straighten up many mathematical manipulations. X.-H. Guo would like to thank TWAS and CBPF for the finacial support. Part of the work is done in CBPF.

\newpage
\vspace{1cm}

\noindent Figure Caption \\

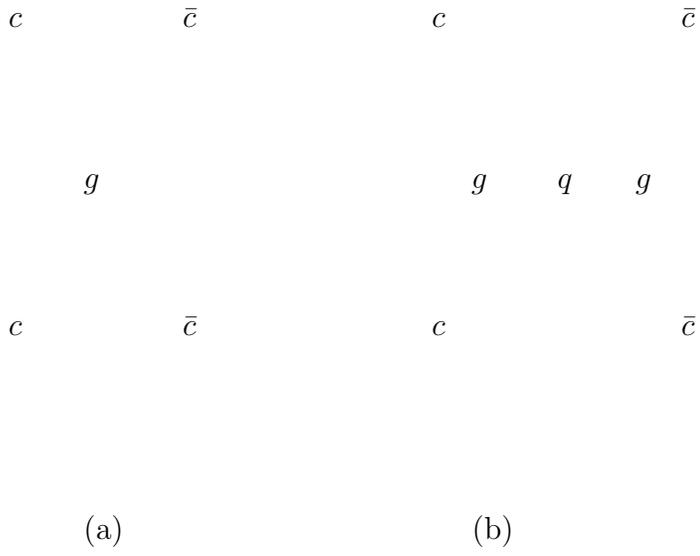
Fig.1.The Feynman diagrams of t-channel $c-\bar c$ scattering where the
charm-quark may deviate from its mass shell a bit.

\newpage
\vspace{3cm}
{\hskip 2cm}$c$ {\hskip 2cm}$\bar{c}$ {\hskip 3cm}$c$ {\hskip 3cm}$\bar{c}$

\vspace{1.2cm}
{\hskip 3cm}$g$ {\hskip 4.8cm}$g$ {\hskip 0.8cm}$q$ {\hskip 0.7cm}$g$

\vspace{1cm}
{\hskip 2cm}$c$ {\hskip 2cm}$\bar{c}$ {\hskip 3cm}$c$ {\hskip 3cm}$\bar{c}$\\

\vspace{1cm}
{\hskip 3cm}(a) {\hskip 4.5cm}(b)\\

\vspace{1cm}
{\hskip 6cm}{\Large \bf Fig.1} 

\end{document}